\documentclass[journal]{IEEEtran}
\usepackage{amsmath,amsfonts}
\usepackage{algorithmic}
\usepackage{algorithm}
\usepackage{array}
\usepackage[caption=false,font=normalsize,labelfont=sf,textfont=sf]{subfig}
\usepackage{textcomp}
\usepackage{stfloats}
\usepackage{url}
\usepackage{verbatim}
\usepackage{graphicx}
\usepackage{cite}
\usepackage{bm}
\usepackage{threeparttable}
\usepackage{booktabs} 
\usepackage[flushleft]{threeparttablex}
\begin{document}

\title{Comparison of FTN-NOFDM and PCS-OFDM for Long-Haul Coherent Optical Communications}

\author{Haide Wang, Ji Zhou, Yongcheng Li, Weiping Liu, Changyuan Yu, Xiangjun Xin, and Liangchuan Li
\thanks{Manuscript received; revised. This work was supported in part by National Key R\&D Program of China under Grant 2023YFB2905700, in part by National Natural Science Foundation of China under Grants 62371207 and 62005102, in part by Young Elite Scientists Sponsorship Program by CAST under Grant 2023QNRC001, in part by the Guangzhou Basic and Applied Basic Research Foundation under Grant 2025A04J5117, and in part by Hong Kong Research Grants Council GRF under Grant 15231923. \it{(Corresponding author: Ji Zhou and Liangchuan Li.)}}

\thanks{Haide Wang is with the School of Cyber Security, Guangdong Polytechnic Normal University, Guangzhou 510665, China.}

\thanks{Ji Zhou, Xiangjun Xin, and Liangchuan Li are with the Aerospace and Informatics Domain, Beijing Institute of Technology, Zhuhai, Zhuhai 519088, China.}

\thanks{Yongcheng Li is with the Jiangsu New Optical Fiber Technology and Communication Network Engineering Research Center and School of Electronic and Information Engineering, Soochow University, Suzhou 215006, China.}%

\thanks{Weiping Liu is with the Department of Electronic Engineering, College of Information Science and Technology, Jinan University, Guangzhou 510632, China.}

\thanks{Changyuan Yu is with the Department of Electronic and Information Engineering, The Hong Kong Polytechnic University, Hong Kong.}
}
\markboth{}%
{Shell \MakeLowercase{\textit{et al.}}: Bare Demo of IEEEtran.cls for IEEE Journals}

\makeatletter
\def\ps@IEEEtitlepagestyle{%
  \def\@oddfoot{\mycopyrightnotice}%
  \def\@evenfoot{}%
}
\def\mycopyrightnotice{%
  {\hfill \footnotesize  \copyright 2025 IEEE\hfill}
}
\makeatother

\maketitle

\begin{abstract}
Single-wavelength 400G coherent optical communications have become a critical solution to meet the explosive traffic demands. However, the single-carrier modulation using low-order modulation formats requires a broader wavelength division multiplexing grid and expands the occupied optical bandwidth. In this paper, we propose the faster-than-Nyquist non-orthogonal frequency division multiplexing (FTN-NOFDM) to improve the spectral efficiency for long-haul coherent optical communications. The subcarrier number is set to eight to enable low-complexity FTN-NOFDM signal generation using a pruned inverse fast Fourier transform and inter-carrier interference (ICI) cancellation. To deal with the conventional timing recovery (TR) failure, a frequency tone-based TR is proposed for FTN-NOFDM. A time-domain multiple-input multiple-output equalizer is designed to update the tap coefficients based on outputs of conventional iterative detection (ID). To further mitigate ICI, a low-density parity check-assisted ID is integrated into the conventional ID module. FTN-NOFDM, probabilistic constellation shaping (PCS)-OFDM, and quadrature phase shift keying-OFDM are experimentally compared in a 400G coherent optical communication system over 11 cascaded 125-GHz wavelength-selective switches (WSSs) and 2000 km transmission. Results show that the FTN-NOFDM exhibits comparable WSS filtering tolerance to PCS-OFDM and superior nonlinearity tolerance, while PCS-OFDM achieves the best bit error ratio performance. 
\end{abstract}


\section{Introduction}
\IEEEPARstart{R}{ecently}, driven by the rapid development of emerging services, such as generative artificial intelligence, computing power networks, and the 5G Advanced, the capacity demand of optical transport networks has surged \cite{pincemin2022end,zuo2023field,ge2023fully}. Single-wavelength 400G coherent optical communications employing quadrature phase shift keying (QPSK) at up to 130 Gbaud have been a critical solution to meet the explosive data traffic demands \cite{pedro2020optical}. Nevertheless, increasing the baud rate inevitably leads to a broader wavelength division multiplexing (WDM) grid, consequently expanding the occupied optical bandwidth. Adopting the higher-order modulation formats like probabilistic constellation shaping (PCS) 16-ary quadrature amplitude modulation (16QAM) could improve the spectral efficiency (SE) for WDM systems under the grid constraints \cite{cho2019probabilistic,arnould2020field, lorences2021pcs}, but compromises nonlinearity tolerance and optical signal-to-noise ratio (OSNR) \cite{lavigne2016400, li2019non, castrillon2020first}. Moreover, the bits of PCS-16QAM have increased due to the redundancy. 

Although single-carrier modulation has dominated in optical transport networks, single-carrier optical communication systems at high baud rates face fundamental limitations, such as the equalization-enhanced phase noise, the demanding parallelism requirements of digital signal processing (DSP), and so forth. In the WDM link, the optical filtering distortions caused by cascading multi-level wavelength-selective switches (WSSs) limit the effective bandwidth of the link, which poses higher requirements for the compensation DSP algorithms of single-carrier modulation. As a result, high baud-rate optical communication systems have moved towards multi-carrier modulation solutions to solve the challenges faced by single-carrier modulation. The multi-carrier modulation mainly includes digital subcarrier multiplexing \cite{sun2020800g,welch2023digital,wang2023fast}, orthogonal frequency division multiplexing (OFDM) \cite{lowery2007fiber, shieh2008coherent, liu2011448, wu2024analysis}, and faster-than-Nyquist non-orthogonal frequency division multiplexing (FTN-NOFDM) \cite{darwazeh2013optical,liu2021faster,ding2021generation}. 

FTN-NOFDM improves the SE by reducing the spacing of the subcarriers \cite{kanaras2009spectrally, zhou2017capacity}. However, FTN-NOFDM coherent optical communication systems suffer from the failure of conventional timing recovery (TR) and the performance penalty caused by inter-carrier interference (ICI). On the one hand, the conventional TR algorithms, such as Gardner and Godard, rely on the Nyquist-frequency components to estimate the timing error, but they are absent in the FTN-NOFDM signal. Therefore, complicated TRs have been proposed for FTN signals, such as the advanced timing phase detection \cite{huang2024interval}, equalizer-aided TR \cite{nakamura2024clock}, and accurate TR loop design \cite{deng2024robust}. On the other hand, the non-orthogonal subcarrier spacing causes ICI, which limits the performance of FTN-NOFDM systems. Advanced detection algorithms, such as sphere decoding, iterative detection (ID), and maximum likelihood detector, are required to cancel the ICI for FTN-NOFDM systems. However, with the increase of the subcarrier number, the computational complexity of the detection algorithms and the signal generation increase significantly.

In our previous work \cite{zhou2021non}, we have proposed the non-orthogonal discrete multi-tone for optical networks toward higher SE. In this paper, we propose the FTN-NOFDM to improve SE for long-haul coherent optical communications. To enable low-complexity FTN-NOFDM signal generation using a pruned inverse fast Fourier transform (IFFT) and ICI cancellation, the subcarrier number is set to eight. A frequency tone-based TR is proposed for FTN-NOFDM. A time-domain (TD) multiple-input multiple-output (MIMO) equalizer is designed that updates the tap coefficients based on outputs of the conventional ID. To further mitigate ICI, a low-density parity check (LDPC)-assisted ID is integrated into the conventional ID module. The main contributions of this paper are as follows:
\begin{itemize}
\item DSP for the FTN-NOFDM signal is introduced, including the pruned IFFT-based signal generation, frequency tone-based TR algorithm, TD-MIMO equalizer, and LDPC-assisted ID for ICI cancellation. 
\item QPSK-NOFDM, PCS16-OFDM, and QPSK-OFDM are experimentally compared in a 400G coherent optical communications over 11 cascaded WSSs and 2000 km transmission.   
\end{itemize}

\begin{figure}[!t]
	\centering
	\includegraphics[width =\linewidth]{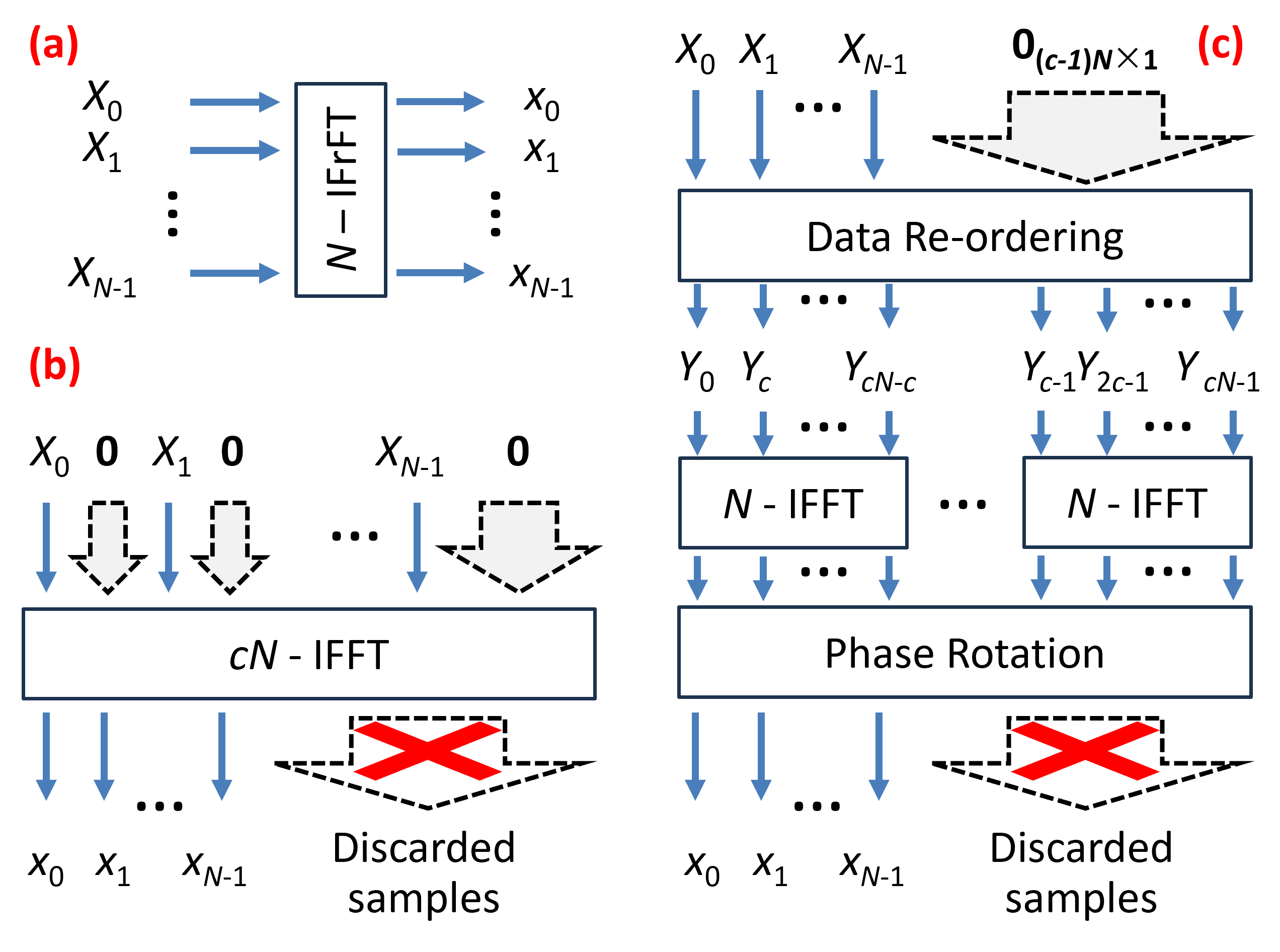}
	\caption{Schematic diagram of the FTN-NOFDM signal generation schemes based on (a) $N$-IFrFT, (b) $cN$-IFFT, and (c) multiple $N$-IFFT.}
	\label{IFFT}
\end{figure}

The rest of the paper is organized as follows. In Section \ref{Principle}, the DSP principle of the FTN-NOFDM system is presented. The experimental setups of the 400G long-haul coherent optical communications are demonstrated in Section \ref{ES}. The experimental results and discussions are given in Section \ref{RESULTS}. Finally, the paper is concluded in Section \ref{CONCLUSION}.

\section{DSP Principle of FTN-NOFDM system}
\label{Principle}
\subsection{FTN-NOFDM signal generation}

\subsubsection{$N$-IFrFT scheme}
As shown in Fig. \ref{IFFT}(a), the time-domain FTN-NOFDM signal with $N$ subcarriers can be generated based on the $N$-point inverse fractional Fourier transform (IFrFT) as \cite{isam2010simple, whatmough2012vlsi}
\begin{equation}
x_n=\frac{1}{\sqrt{N}} \sum_{k=0}^{N-1} X_k \cdot e^{j\frac{ 2 \pi nk\alpha}{N}}
\end{equation}
where $X_k$ is the frequency-domain FTN-NOFDM signal. The compression factor $\alpha$ can be decomposed as a rational fraction $b/c$ ($c > b$). 

\begin{figure}[!t]
	\centering
	\includegraphics[width =\linewidth]{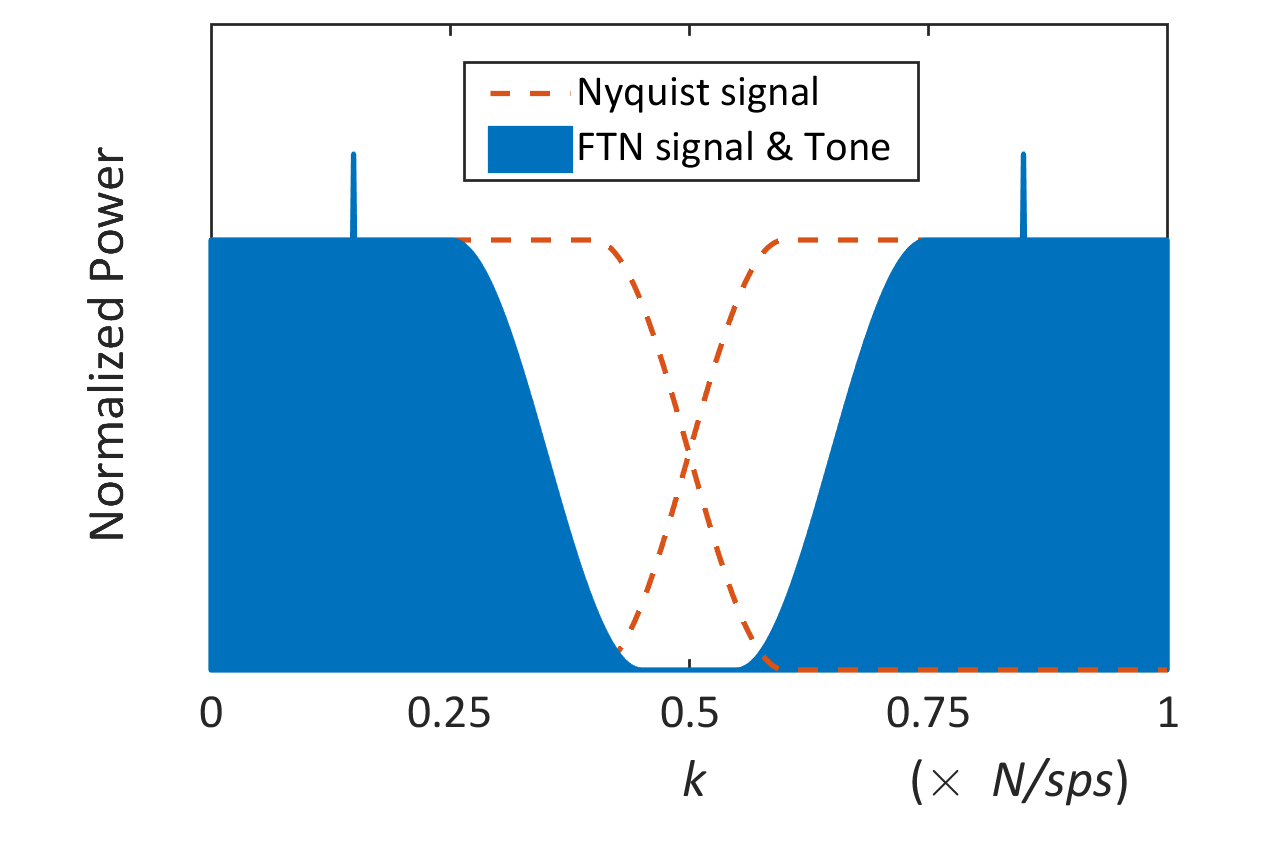}
	\caption{Schematic diagram of the spectrum of the Nyquist signal and the FTN signal with tone.}
	\label{spectrum}
\end{figure}

\subsubsection{$cN$-IFFT scheme}
As shown in Fig. \ref{IFFT}(b), the FTN-NOFDM signal can also be generated by the $cN$-point IFFT as
\begin{equation}
x_n=\frac{1}{\sqrt{N}} \sum_{k=0}^{N-1} X_k \cdot e^{j\frac{ 2 \pi n(kb)}{cN}}=\frac{1}{\sqrt{N}} \sum_{k^\prime=0}^{cN-1} Y_{k^\prime} \cdot e^{j\frac{ 2 \pi n{k^\prime}}{cN}}
\label{cN_IFFT_scheme}
\end{equation}
where the input signal $Y_{k^\prime}$ can be expressed as
\begin{equation}
Y_{k^\prime}=\left\{
\begin{array}{cc}
X_{k^\prime/b}, &k^\prime = 0,b,\cdots,(N-1)b\\ 0, &\text{other}
\end{array}\right..
\end{equation}

\subsubsection{Multiple $N$-IFFT scheme}
To implement FTN-NOFDM signal generation more efficiently, the multiple $N$-IFFT can also be used as shown in Fig. \ref{IFFT}(c). The signal $X_k$ appended $(c-1)N$ zeros are first re-ordered as the $Y_{i+mc}$. By replacing $k^\prime$ in Eq. (\ref{cN_IFFT_scheme}) with $i+mc$, the output signal of the multiple $N$-IFFT is phase rotated and summed up to generate the FTN-NOFDM signal as
\begin{equation}
\begin{aligned}
x_n&=\frac{1}{\sqrt{N}} \sum_{i=0}^{c-1}\sum_{m=0}^{N-1} Y_{i+m c} e^{\frac{j 2 \pi n (i+m c)}{c N}} \\&=\frac{1}{\sqrt{N}}\sum_{i=0}^{c-1}e^{\frac{j 2 \pi n i}{c N}} \sum_{m=0}^{N-1}Y_{i+m c} e^{\frac{j 2 \pi n m}{N}}.
\end{aligned}
\end{equation}

\subsubsection{Analysis of computational complexity}
The scheme for FTN-NOFDM signal generation should be chosen according to the $N$, $b$, and $c$. $N^2$ complex multiplications and $N(N-1)$ additions are required by the $N$-IFrFT scheme. The upper bound of multiplications for the $cN$-IFFT, and multiple $N$-IFFT schemes are $1/2\cdot cN \log_2cN$ and $1/2\cdot cN \log_2N + cN$, respectively. The upper bound of additions for the $cN$-IFFT, and multiple $N$-IFFT schemes are $cN \log_2cN$ and $cN \log_2N + (c-1)N$, respectively. The subcarrier number $ N$ should be kept small to reduce the overall computational complexity of FTN-NOFDM signal generation and ICI cancellation. When $c$ is close to $N$, the computational complexity of $N$-IFrFT will be lower than that of multiple $N$-IFFT due to the phase rotation operation. Due to the zero input and the discarded output at the $cN$-IFFT, its trellis can be pruned to remove the redundant calculation as \cite{markel1971fft, sreenivas1979fft, sorensen1993efficient}
\begin{equation}
TS=\left\{
\begin{array}{cc}
\left[2 Q-I-O-2\left(1-2^{O-Q}\right)\right]/Q, &I+O \ge Q \\[1ex] \left[Q-I-2^{I+1-Q}\left(1-2^{O}\right)\right]/Q, &I+O < Q
\end{array}\right.
\end{equation}
where $Q=\log_2cN$. The number of non-zero inputs and desired output signal is $2^I$ and $2^O$, respectively.

\begin{figure}[!t]
	\centering
	\includegraphics[width =\linewidth]{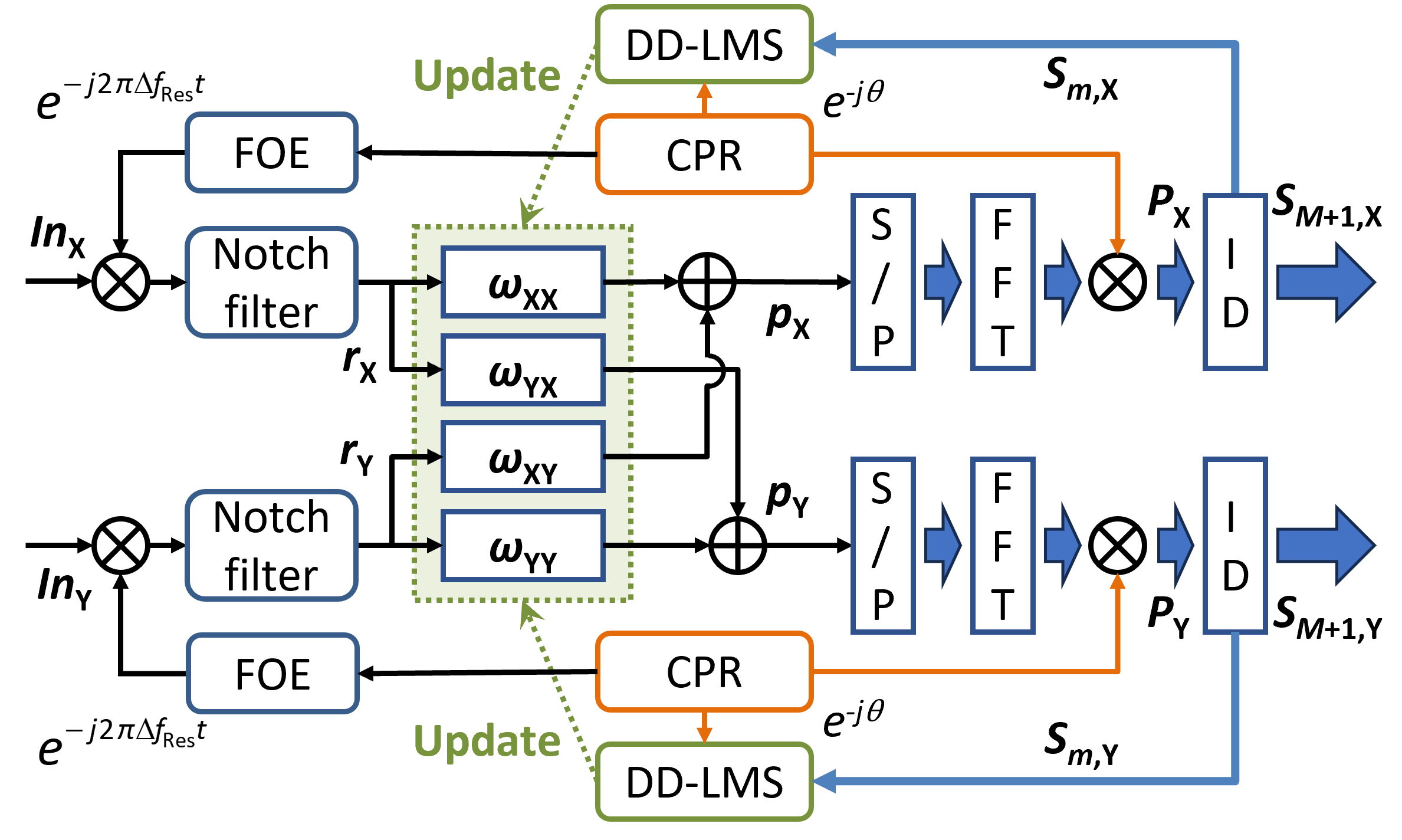}
	\caption{Schematic diagrams of the TD-MIMO equalizer combined with the conventional ID algorithm for the FTN-NOFDM signal.}
	\label{FFE}
\end{figure}

\subsection{Frequency tone-based timing recovery for FTN signal}
The schematic diagram of the spectrum of the Nyquist signal and the FTN signal with tone is shown in Fig. \ref{spectrum}. The two parts of these two signals are shifted by $(N-N/sps)$, where $N$ is the discrete Fourier transform size and $sps$ is the oversampling rate. The overlap of the Nyquist-frequency components of the Nyquist signal is used to estimate the sampling phase offset by conventional TR algorithms \cite{wang2023non}. However, for the FTN signal, there is no overlap of the Nyquist-frequency components. A frequency tone at $1/K$ baud rate is used to realize a low-complexity TR for the FTN signal, which estimates the sampling phase offset as
\begin{equation}
\begin{aligned}
e&=\sum_{k=F_{R_{\text{s}}/K}-l}^{F_{R_{\text{s}}/K}+l-1}\operatorname{Im}\left[X(k) \cdot X^*(N-2 N / sps / K+k)\right] \\
&\propto \sin \left(\frac{4 \pi \tau}{sps K}\right)  
\end{aligned}
\end{equation}
where $\operatorname{Im}\left(\cdot\right)$ denotes the imaginary part of a complex value. $F_{R_{\text{s}}/K} = N/sps/K$ and  $2l$ frequency points around the frequency tones at $R_{\text{s}}/K$ are also used. $\left(\cdot\right)^*$ is the complex conjugate. After TR, the frequency tones can be filtered by a notch filter \cite{ingle2011digital}, of which the transfer function can be expressed as
\begin{equation}
H(z)=\frac{1-2 \cos \theta \cdot z^{-1}+z^{-2}}{1-2 r_0 \cdot \cos \theta \cdot z^{-1}+r_0^2 \cdot z^{-2}}
\end{equation}
where $\theta$ is the notch frequency. $r_0$ represents the radius of the pole, which ranges from zero to one.

\subsection{TD-MIMO Equalizer for the FTN-NOFDM signal}
The time-domain equalizer would be more efficient for the FTN-NOFDM signal with a relatively small subcarrier number $N$. Fig. \ref{FFE}(a) shows the schematic diagram of the proposed $2\times 2$ TD-MIMO equalizer combined with the conventional ID algorithm for the FTN-NOFDM signal. After compensating for the frequency offset and filtering the tone sequence using a notch filter, the output of the TD-MIMO equalizer can be calculated as
\begin{equation}
p_\mathrm{X}(n) = \bm{r}_\mathrm{X}(n)\times \boldsymbol{\omega}_\mathrm{XX}^\mathrm{H} + \bm{r}_\mathrm{Y}(n)\times \boldsymbol{\omega}_\mathrm{XY}^\mathrm{H}
\end{equation}
and
\begin{equation}
p_\mathrm{Y}(n) = \bm{r}_\mathrm{Y}(n)\times \boldsymbol{\omega}_\mathrm{YY}^\mathrm{H} + \bm{r}_\mathrm{X}(n)\times \boldsymbol{\omega}_\mathrm{{YX}}^\mathrm{H}
\end{equation}
where $(\cdot)^ {\mathrm{H}}$ denotes the conjugate transpose operation. $\boldsymbol{\omega}_{\mathrm{{XX}}}$, $\boldsymbol{\omega}_\mathrm{{XY}}$, $\boldsymbol{\omega}_\mathrm{{YX}}$, and $\boldsymbol{\omega}_\mathrm{{YY}}$ are the tap coefficient vectors with a length of $L$. $\bm{r}_{\mathrm{X}}(n) = \left[r_{\mathrm{X}}(n-L+1), \dots, r_{\mathrm{X}}(n)\right]$ and $\bm{r}_{\mathrm{Y}}(n) = \left[r_{\mathrm{Y}}(n-L+1), \dots, r_{\mathrm{Y}}(n)\right]$ are the received signal vectors at the X and Y polarization, respectively.

\begin{figure}[!t]
	\centering
	\includegraphics[width =\linewidth]{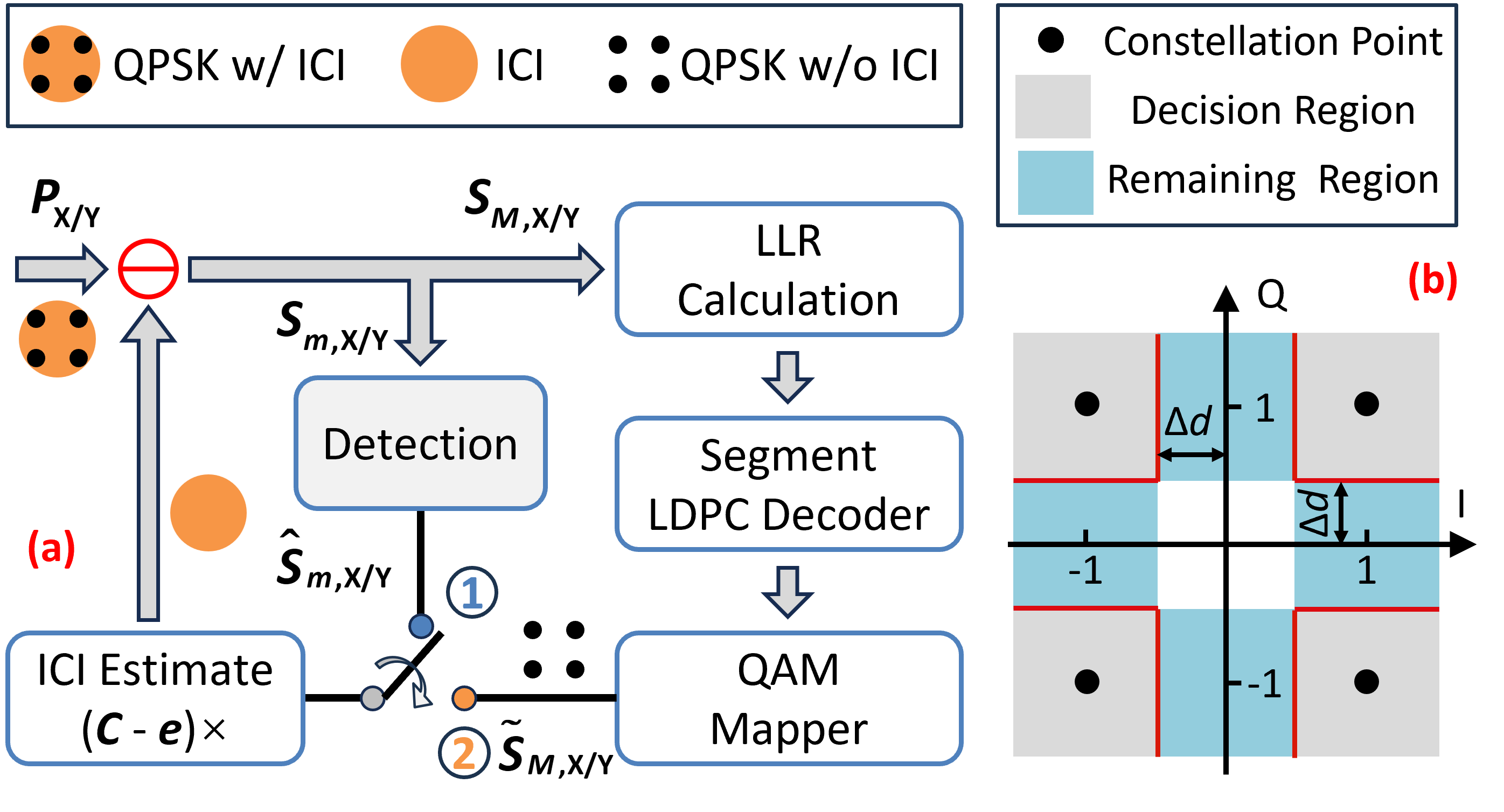}
	\caption{(a) Schematic diagram of the LDPC-assisted ID for ICI cancellation. (b) The schematic diagram of the detection of the conventional ID algorithm.}
	\label{ID}
\end{figure}

After serial-to-parallel (S/P) conversion, the equalized FTN-NOFDM signal $\bm{p}_{\mathrm{X/Y}}(n)$ is then transformed to the frequency-domain signal by a pruned FFT. However, the MIMO equalizer can not deal with ICI, which causes errors in the update of tap coefficients. Thus, the conventional ID algorithm is used to obtain the signal $S_\mathrm{X/Y}(k)$ without ICI before the feedback for the update of tap coefficients.

The tap coefficients of the TD-MIMO equalizer for the FTN-NOFDM signal are updated by the decision-directed least-mean square (DD-LMS) algorithm as
\begin{equation}
\bm{\omega}_{\bullet \mathrm{X}} = \bm{\omega}_{\bullet \mathrm{X}} + \mu e^{-j\theta}\sum_{n=0}^{N-1}\left[\bm{d}_{m,\mathrm{X}}(n) - \bm{s}_{m,\mathrm{X}}(n)\right]^{*}\times \bm{r_\mathrm{X}}(n)
\end{equation}
and
\begin{equation}
\bm{\omega_{\bullet \mathrm{Y}}} = \bm{\omega_{\bullet \mathrm{Y}}} + \mu e^{-j\theta}\sum_{n=0}^{N-1}\left[\bm{d}_{m,\mathrm{Y}}(n) - \bm{s}_{m,\mathrm{Y}}(n)\right]^{*}\times \bm{r_\mathrm{Y}}(n)
\end{equation}
where the subscript $(\cdot)_\bullet$ denotes X or Y. The $(\cdot)^*$ is the conjugation operation. $\mu$ is the step size and $\theta$ is the estimated phase noise. $\bm{s}_{m,\mathrm{X/Y}}(n)$ is the IFFT of output signal $\bm{S}_{m,\mathrm{X/Y}}(k)$ for ID algorithm at the $m$-th iteration. $\bm{d}_{m,\mathrm{X/Y}}(n)$ is the IFFT of $\bm{S}_{m,\mathrm{X/Y}}(k)$ after hard-decision.

\begin{figure*}[!t]
	\centering
	\includegraphics[width =\linewidth]{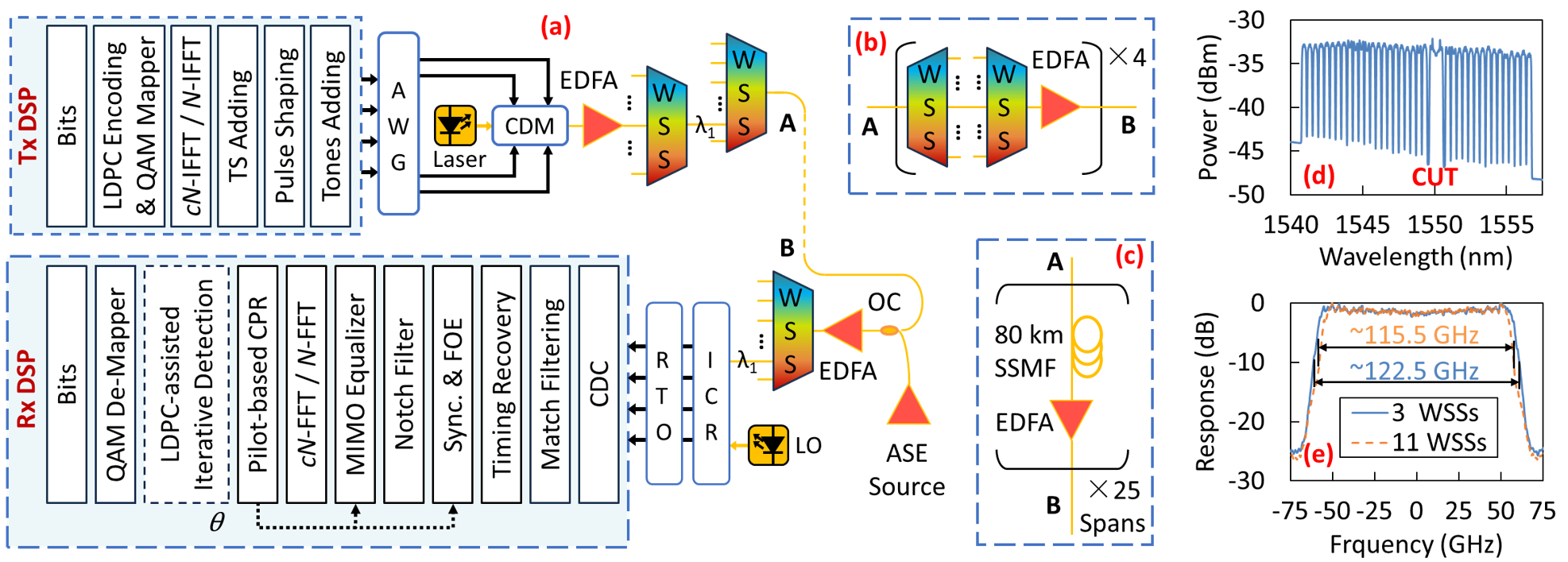}
	\caption{(a) Experimental setups for the 400G long-haul coherent optical communication system with three WSSs over OBtB transmission. The link configurations of (b) 8 cascaded WSSs transmission link and (c) 2000 km SSMF transmission link. (d) The measured optical spectrum of the WDM system. (e) The measured frequency response of the channel under test (CUT) with three WSSs over OBtB and the 8 cascaded WSS transmission link.}
	\label{EX}
\end{figure*}

\subsection{LDPC-assisted iterative detection for ICI cancellation}
Fig. \ref{ID}(a) shows the schematic diagram of the proposed LDPC-assisted ID algorithm for ICI cancellation. When the switch is on branch 1, it is the conventional ID algorithm \cite{huang2016experimental}. At the $m$-th iteration, the decision threshold $\Delta d$ contracts as
\begin{equation}
\Delta d = 1 - m/M
\end{equation}
where $M$ denotes the total iteration number. The schematic diagram of the detection of the conventional ID algorithm is shown in Fig. \ref{ID}(b) for the QPSK signal. The symbols in the decision region are made for the hard decision, while those in the remaining region are unchanged. After ICI cancellation, the output at the $m$-th iteration can be expressed as
\begin{equation}
\bm{S}_{m,\mathrm{X/Y}} = \bm{P}_\mathrm{X/Y} - (\bm{C}- \bm{e}) \times \hat{\bm{S}}_{m-1,\mathrm{X/Y}}
\end{equation}
where $\bm{P}_{\mathrm{X/Y}}$ is the equalized FTN-NOFDM signal in the frequency domain. $\bm{e}$ and $\bm{C}$ denote the $N\times N$ unit matrix and the interference matrix, respectively. The output signal $\bm{S}_{m,\mathrm{X/Y}}$ with fewer than $M$ iterations can be employed to update the tap coefficients of the MIMO equalizer using the DD-LMS algorithm to reduce the loop delay. 

Although the conventional ID algorithm has relatively lower complexity, especially when the subcarrier number and the total iteration number are small, the ability to cancel ICI is limited. Therefore, an LDPC-assisted ID is integrated into the conventional ID module to improve the bit error ratio (BER) performance of the QPSK-NOFDM signal. When the switch is on branch 2 at the $(M+1)$-th iteration, it becomes the LDPC-assisted ID for ICI cancellation as shown in Fig. \ref{ID}(a). Firstly, the log-likelihood ratio (LLR) of the signal at the $M$-th iteration is calculated for the segment LDPC decoder \cite{mo2023simplified}. Part of the iterations of the sum-product algorithm are used for LDPC-assisted ID, while the remaining iterations are reserved for the signal after ICI cancellation. Then the output of LDPC decoding is mapped to the  QPSK symbols $\tilde{\bm{S}}_{M, \mathrm{X/Y}}$. A more accurate ICI of the X or Y polarization in the frequency domain can be estimated as 
\begin{equation}
\bm{S}_{M+1,\mathrm{X/Y}} = \bm{P}_\mathrm{X/Y} - (\bm{C}- \bm{e}) \times \tilde{\bm{S}}_{M,\mathrm{X/Y}}
\end{equation}
Since the proposed LDPC-assisted ID algorithm reuses the ICI estimation module, no additional FFT/IFFT is required \cite{guo2021ldpc}.

\section{Experimental setups of 400G system}
\label{ES}
The experimental setups for the 400G long-haul coherent optical communication system with three WSSs over optical back-to-back (OBtB) transmission are shown in Fig. \ref{EX}(a). At the transmitter (Tx), the bit sequence is encoded by LDPC with an overhead of 20\% and mapped into the QPSK/PCS-16QAM symbols. A $ cN$-IFFT generates the FTN symbols, or the OFDM symbols are generated by the $N$-FFT, where $N=8$. The compression factor $\alpha$ of the QPSK-NOFDM signal is set to 0.875 (i.e., $b/c=7/8$). The training sequence (TS) for frame synchronization, frequency offset estimation (FOE), and one pilot every four FTN-NOFDM/OFDM symbols for carrier phase recovery (CPR) is added. Then the signal is pulse-shaped by a root-raised cosine (RRC) filter with a roll-off factor of 0.01. Then tones at 1/4 baud rate are added for TR. The signal-to-tone power ratio is set to 13 dB. 

An arbitrary waveform generator (AWG) operating at 175 GSa/s transfers the digital signal to the analog signal. The baud rates of the QPSK-OFDM and QPSK-NOFDM signals are both 130 Gbaud, while that of the PCS16-OFDM with the entropy of about 2.6 bits/symbol is 113.75 (i.e., 130$\times$0.875) Gbaud. The net SE is about 3.32 bit/s/Hz. Then the signal is modulated on an optical carrier with a wavelength $\lambda_1$ of $\sim 1550.1$ nm by a coherent driver modulator (CDM). The optical signal is amplified by an Erbium-doped fiber amplifier (EDFA) and passes through two cascaded WSSs. Another 37 WDM channels with 50 GHz bandwidth are multiplexed through the second WSS, while the first WSS is reserved for other WDM channels.

\begin{figure}[!t]
	\centering
	\includegraphics[width =\linewidth]{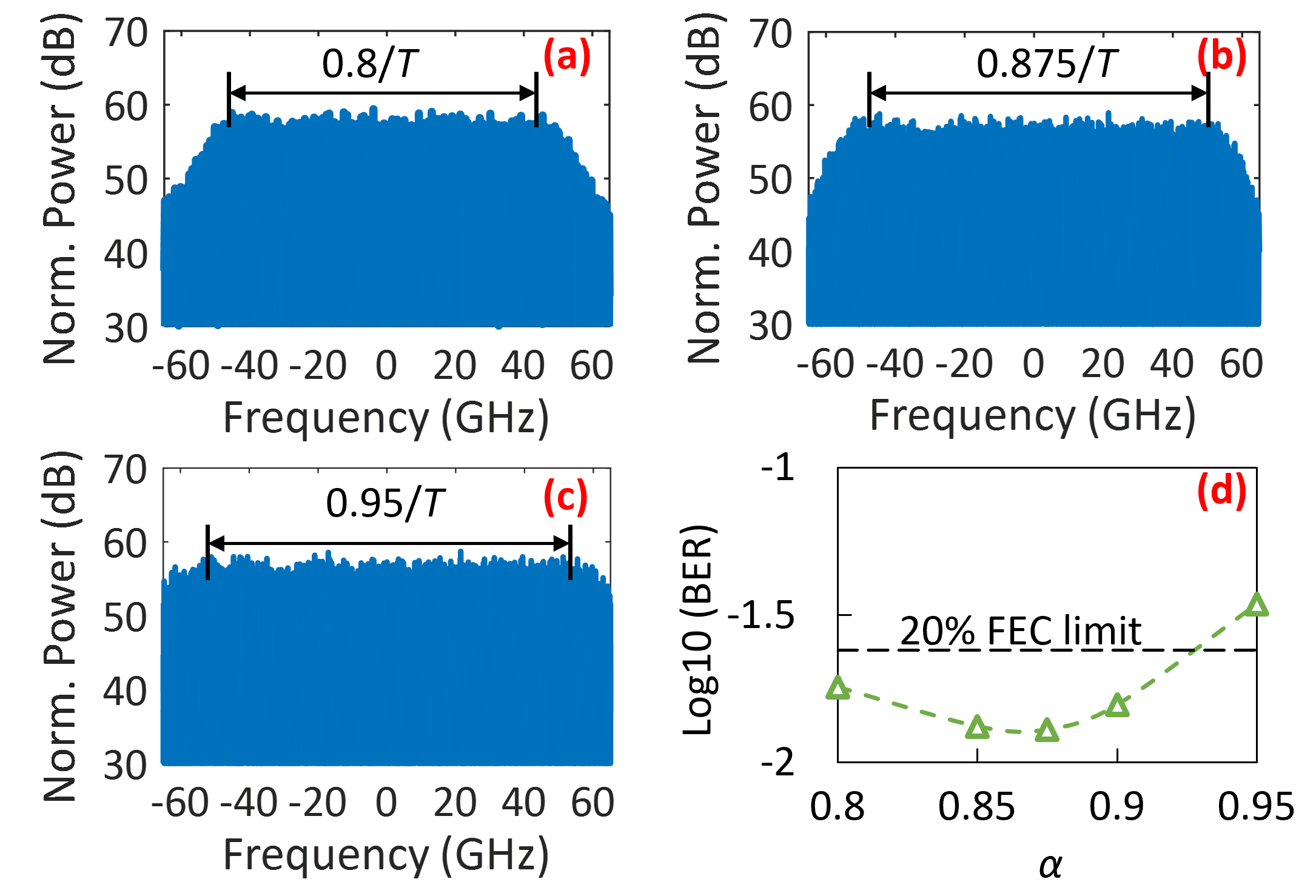}
	\caption{Spectrum of the eight-subcarrier FTN-NOFDM signal with a compression factor $\alpha$ of (a) 0.8, (b) 0.875, and (c) 0.95. (d) BER of FTN-NOFDM signal versus the compression factor $\alpha$ at the OSNR of 20 dB.}
	\label{alpha}
\end{figure}

There are three link configurations, including the OBtB transmission link, the 8 cascaded WSSs transmission link, and the 2000 km standard single-mode fiber (SSMF) transmission link. Different link configurations can be adopted by connecting points A and B in Fig. \ref{EX}(a). At the 8 cascaded WSSs transmission link as shown in Fig. \ref{EX}(b), there are four groups composed of two WSSs and an EDFA. The 2000 km transmission link consists of 25 spans as shown in Fig. \ref{EX}(c), each consisting of 80 km SSMF and an EDFA. Fig. \ref{EX}(d) shows the measured optical spectrum of the WDM system. All the WSSs for the channel under test (CUT) at the wavelength of 1550.1 nm are set to a 125 GHz grid and 50 GHz grids for other WDM channels. The measured frequency response of the CUT with three WSSs over OBtB and the 8 cascaded WSSs transmission link is shown in Fig. \ref{EX}(e). The 10 dB bandwidth is about 122.5 GHz and 115.5 GHz, respectively.

At the receiver (Rx), the amplified spontaneous emission (ASE) noise generated by an EDFA is coupled with the WDM signals by an optical coupler (OC). After being amplified by an EDFA, the CUT is selected by a WSS. An external cavity laser with a linewidth of less than 100 kHz is used as the local oscillator (LO). An integrated coherent receiver (ICR) transfers the optical signal to the analog signal. Then, a real-time oscilloscope (RTO) operating at 256 GSa/s implements the analog-to-digital conversion for the signal. Finally, the Rx DSP recovers the received signal as shown in Fig. \ref{EX}(b). 

Chromatic dispersion compensation (CDC) is first implemented. Then the signal is matched-filtered by an RRC filter. After the frequency tone-based TR, synchronization (sync.), and FOE, the tone is filtered by a notch filter. Then follow the $2\times 2$ TD-MIMO equalizer and the pilot-based CPR. The residual frequency offset $\Delta f_{\text{Res}}$ is estimated by the slope of the estimated phase noise $\theta$. LDPC-assisted ID with an iteration number of 6 is employed to improve the BER performance for the QPSK-NOFDM signal. After 3 iterations of conventional ID with 5 (i.e., $M=5$) total iterations, the output is used for the DD-LMS algorithm to update tap coefficients of the MIMO equalizer. Finally, the BER is calculated.

\begin{figure}[!t]
	\centering
	\includegraphics[width =\linewidth]{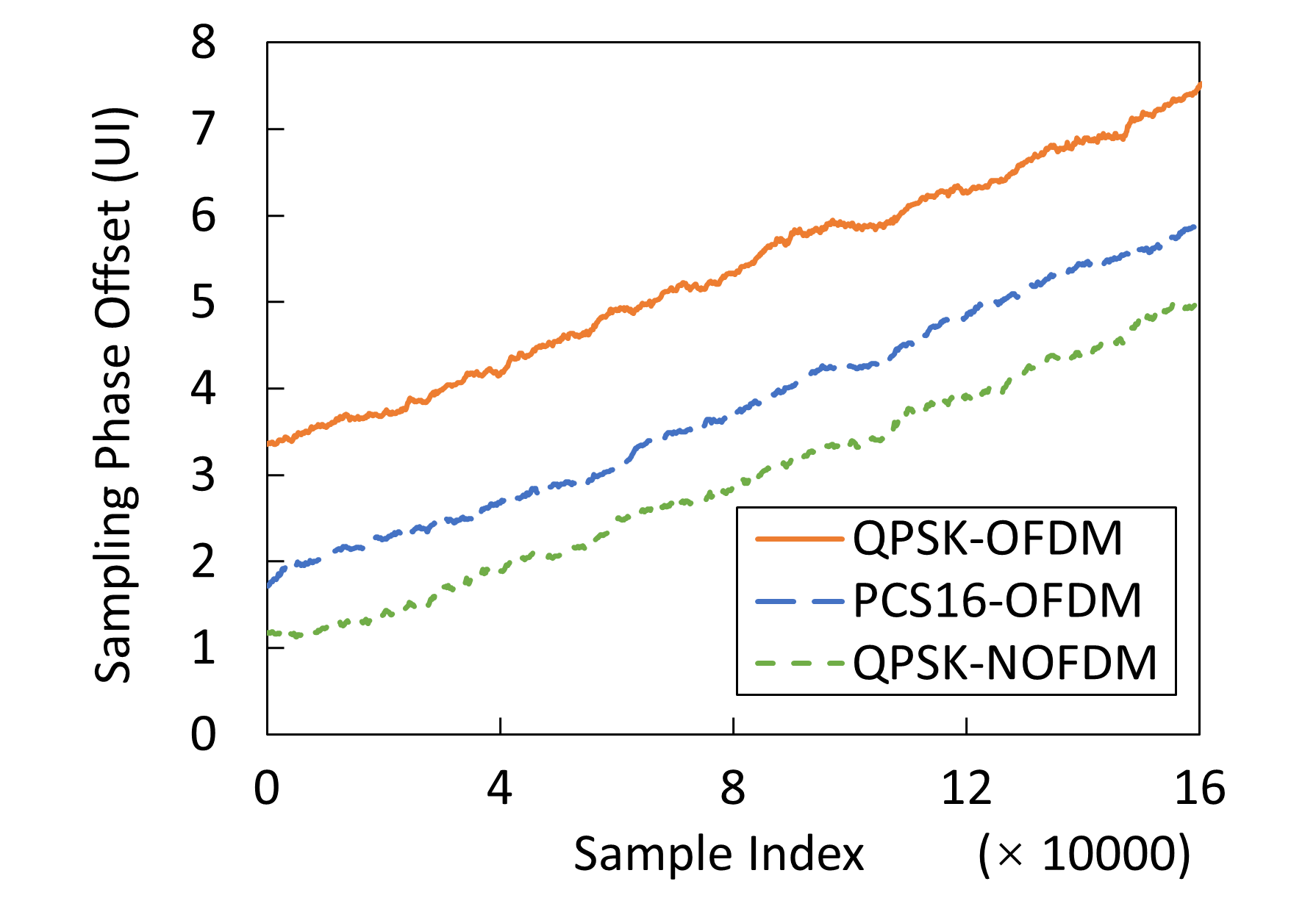}
	\caption{Estimated sampling phase offset by the frequency tone-based TR algorithm of the QPSK-OFDM, PCS16-OFDM, and QPSK-NOFDM signals.}
	\label{SPO}
\end{figure}

\section{Experimental results and discussions}
\label{RESULTS}
The spectrum of the eight-subcarrier FTN-NOFDM signal with a compression factor $\alpha$ of 0.8, 0.875, and 0.95 is shown in Figs. \ref{alpha}(a)-(c), respectively. As the compression factor $\alpha$ decreases, the bandwidth of the FTN-NOFDM signal becomes $\alpha/T$, where $T$ is the duration of the FTN-NOFDM signal. The reduced bandwidth makes the FTN-NOFDM signal resilient to the WSS filtering. Fig. \ref{alpha}(d) shows the BER versus the compression factor $\alpha$ at the OSNR of 20 dB. As the compression ratio $\alpha$ decreases, the bandwidth of the FTN-NOFDM signal narrows, which is beneficial to improve its WSS filtering tolerance. Nevertheless, the influence of ICI becomes more pronounced. Thus, BER decreases and then increases as the compression ratio $\alpha$ decreases. The optimal compression ratio $\alpha$ is about 0.875 (i.e., $b/c=7/8$). Consequently, the FTN-NOFDM signal generation based on the pruned $cN$-IFFT scheme with 56 complex multiplications offers the lowest computational complexity compared to other schemes.

Fig. \ref{SPO} shows the estimated sampling phase offset by the frequency tone-based TR algorithm of the QPSK-OFDM, PCS16-OFDM, and QPSK-NOFDM signals. Due to the slight deviation of the clock frequency of the transceiver, the sampling phase offset of the received signal is linear with the sample index. In addition, the initial sampling phase offset is different for the three signals, but it does not affect the estimate of the clock frequency offset. The sampling phase offset slopes of the three signals estimated by the proposed TR algorithm based on frequency tone are similar, which indicates that there is a clock frequency offset. Therefore, the proposed frequency tone-based TR algorithm can effectively estimate the timing for the FTN-NOFDM signal.  

\begin{figure}[!t]
	\centering
	\includegraphics[width =\linewidth]{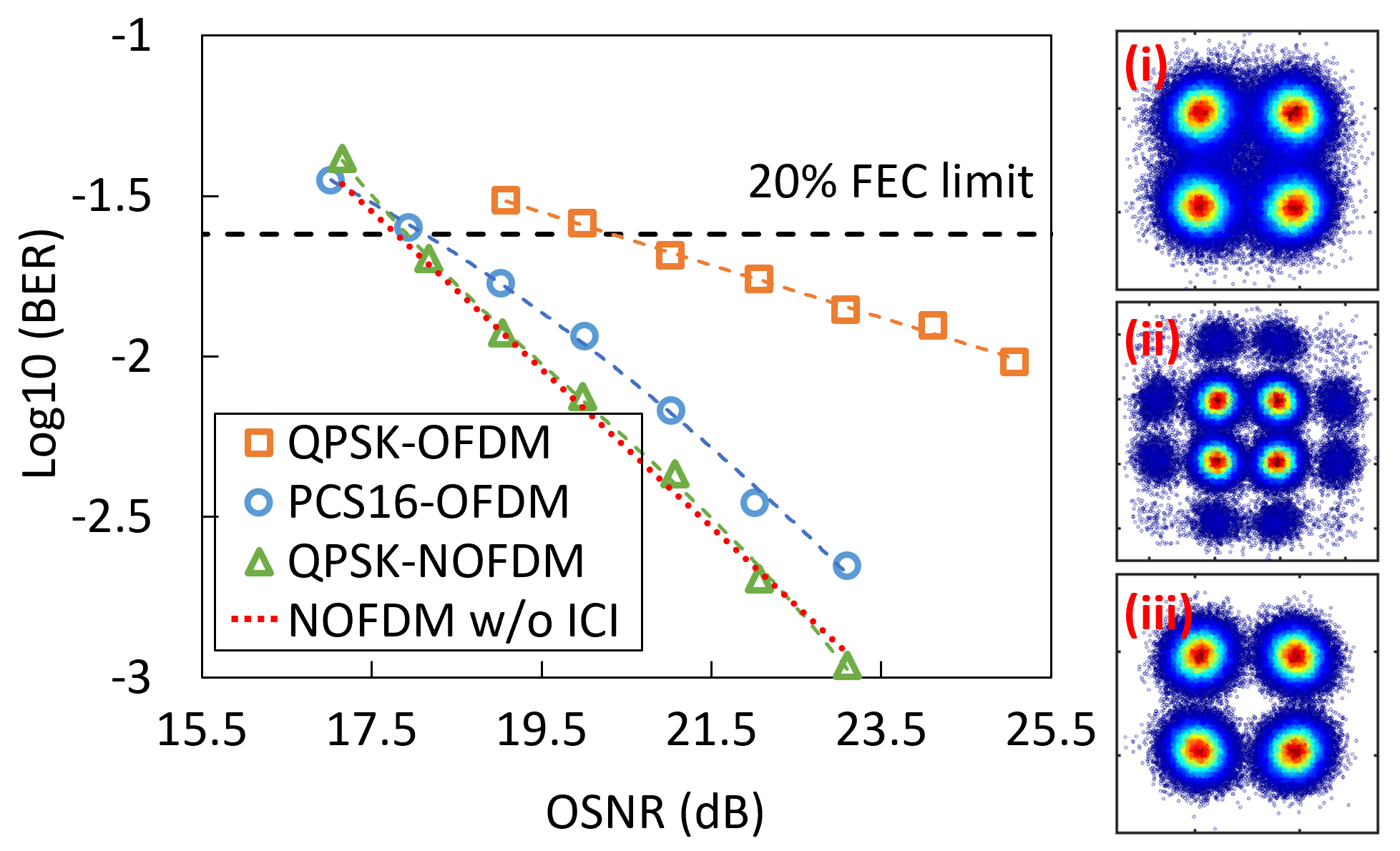}
	\caption{BER versus OSNR of the 400G coherent optical system with three WSSs over OBtB transmission. Insets are the constellation points of (i) QPSK-OFDM, (ii) PCS16-OFDM, and (iii) QPSK-NOFDM signals.}
	\label{BTB_BER}
\end{figure}

\begin{figure}[!t]
	\centering
	\includegraphics[width =\linewidth]{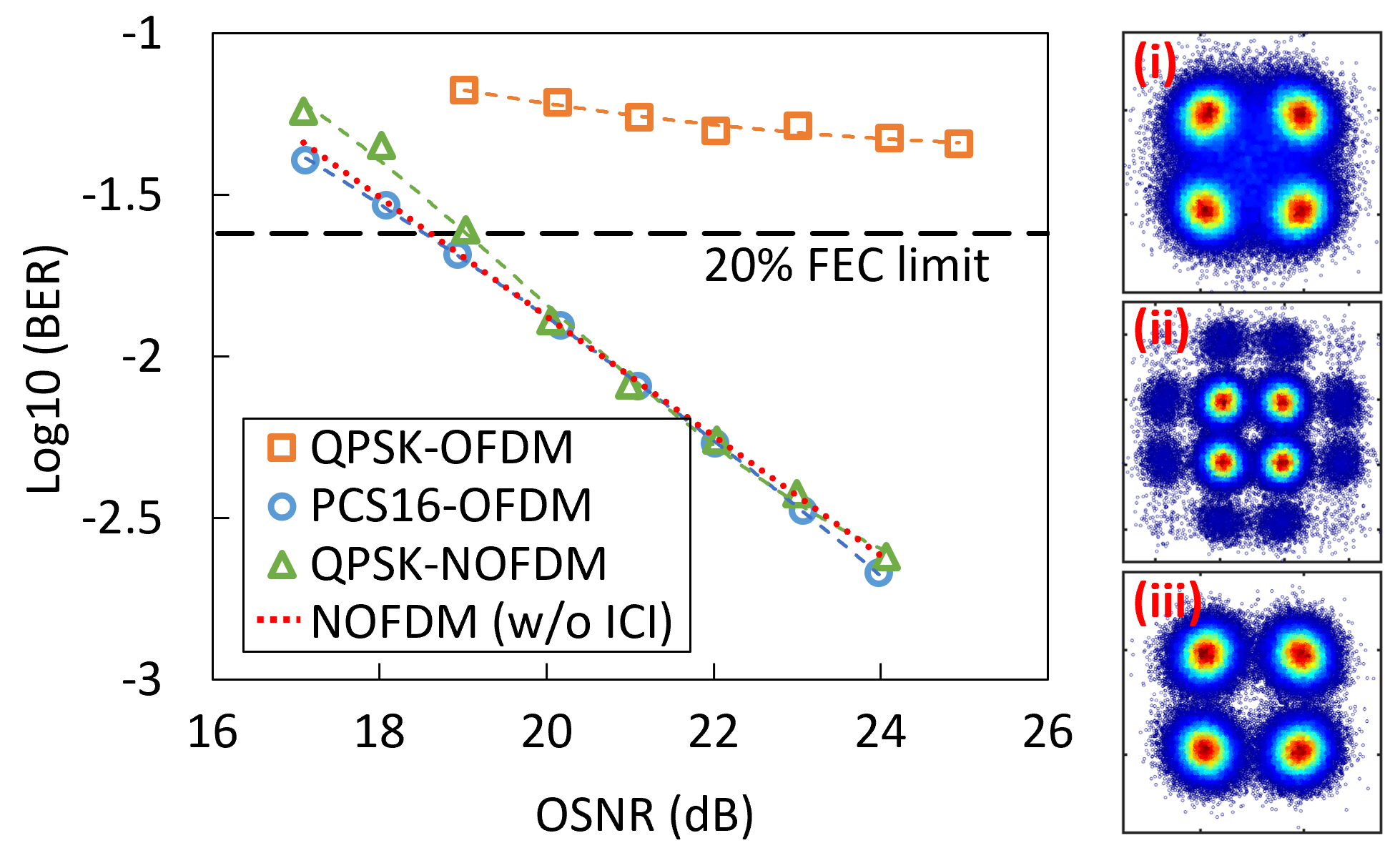}
	\caption{BER versus OSNR of the 400G coherent optical system over 8 cascaded WSSs transmission. Insets are the constellation points of (i) QPSK-OFDM, (ii) PCS16-OFDM, and (iii) QPSK-NOFDM signals.}
	\label{WSS_BER}
\end{figure}

BER versus OSNR of the 400G coherent optical system over OBtB transmission is shown in Fig. \ref{BTB_BER}. Insets are the constellation points of (i) QPSK-OFDM, (ii) PCS16-OFDM, and (iii) QPSK-NOFDM signals at the OSNR of 23 dB. PCS-OFDM and FTN-NOFDM signals can reach the 20\% forward error correction (FEC) limit at an OSNR of about 18 dB and 17.9 dB, respectively. The red dotted line represents the FTN-NOFDM without ICI, which is the upper boundary of the LDPC-assisted ICI. When the OSNR is higher than about 18 dB, the FTN-NOFDM signal can obtain a lower BER than the PCS-OFDM signal. There is about a 2.5 dB OSNR penalty in the QPSK-OFDM at the 20\% FEC limit. From the constellation diagrams, it can be seen that compared with the PCS16-OFDM and QPSK-NOFDM signals, the QPSK-OFDM signal suffers from more severe WSS filtering caused by the three WSSs with a 125-GHz grid. Thus, the QPSK-OFDM signal has a much higher BER than that of the PCS16-OFDM and FTN-NOFDM signals.

BER versus OSNR of the 400G coherent optical system over 8 cascaded WSSs transmission is shown in Fig. \ref{WSS_BER}. Insets are the constellation points of (i) QPSK-OFDM, (ii) PCS16-OFDM, and (iii) QPSK-NOFDM signals at the OSNR of 24 dB. Since the serious WSS filtering limits the BER performance of the subcarriers at high frequencies, the BER of the QPSK-OFDM can not reach the 20\% FEC limit. Owing to being less affected by the WSS filtering, the BER of the QPSK-NOFDM signal is similar to that of the PCS16-OFDM when the OSNR is higher than about 20 dB. Due to the relatively small number of subcarriers, the FTN signal exhibits a large roll-off in the spectrum, resulting in its lower tolerance to the WSS filtering than the lower baud-rate PCS-OFDM signal. When OSNR is lower than 19 dB, the LDPC-assisted ID may not obtain good enough LDPC decoding performance for ICI estimate, resulting in a higher BER of the FTN-NOFDM signal compared to the PCS-OFDM signal. It deviates from the red dotted line of the FTN-NOFDM without ICI. It's expected that the tolerance of the FTN-NOFDM signal to the WSS filtering can be further improved by increasing the subcarrier number, but the computational complexity of the FTN-NOFDM signal generation and ID algorithm will be increased.

\begin{figure}[!t]
	\centering
	\includegraphics[width =\linewidth]{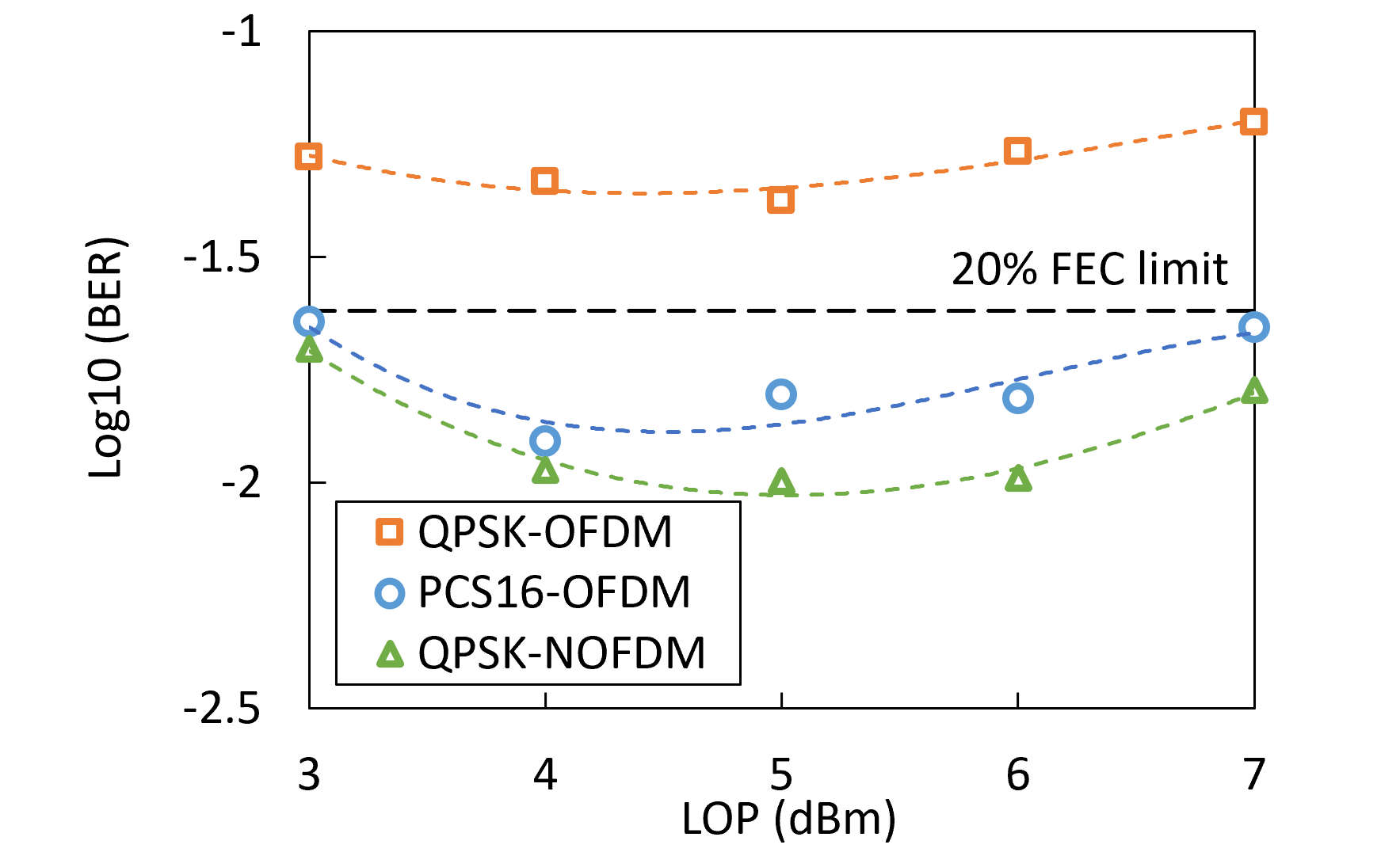}
	\caption{BER of QPSK-OFDM, PCS16-OFDM, and QPSK-NOFDM signals versus LOP of the 400G coherent optical system over 2000 km SSMF transmission.}
	\label{LOP}
\end{figure}

\begin{figure}[!t]
	\centering
	\includegraphics[width =\linewidth]{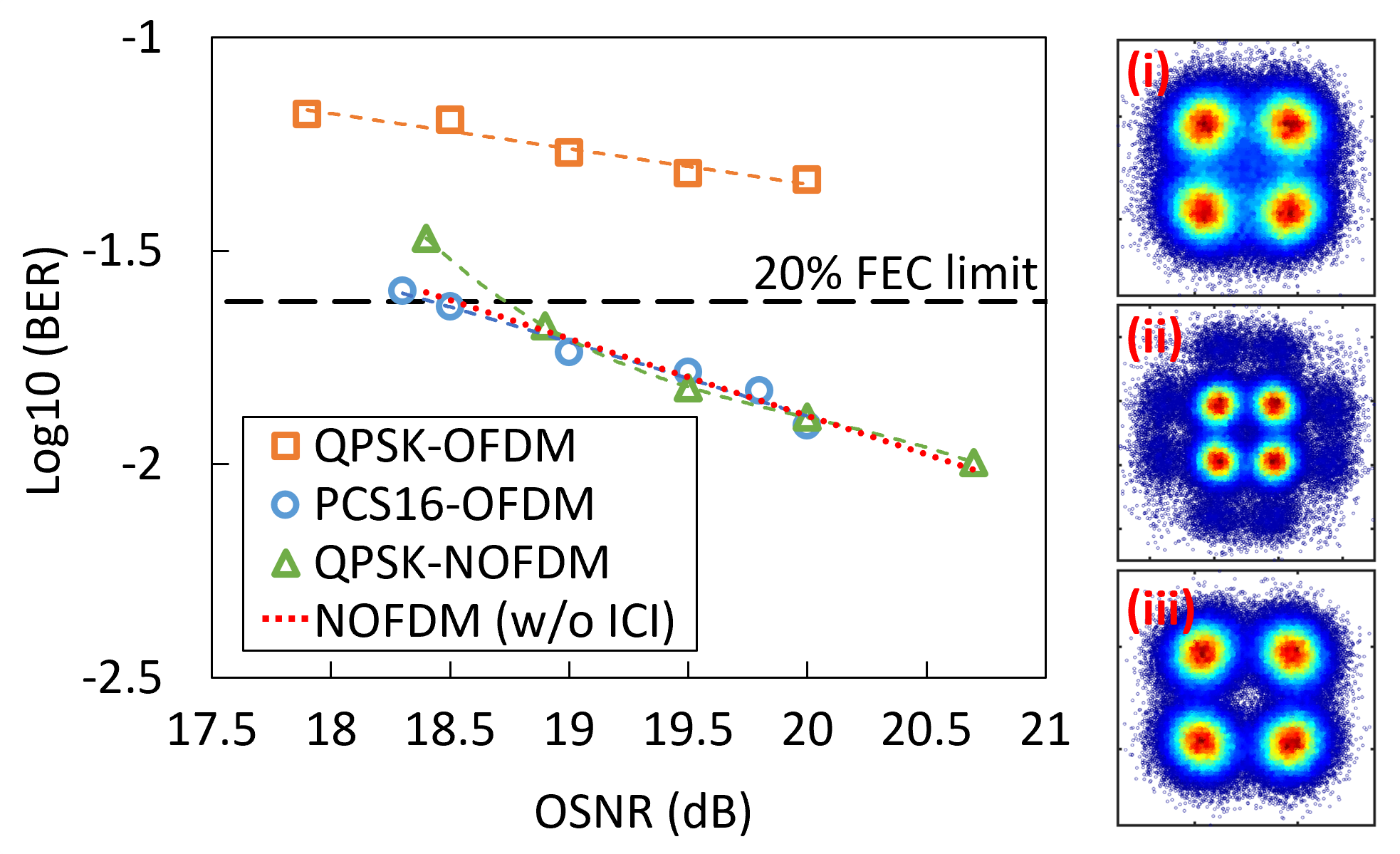}
	\caption{BER versus OSNR of the 400G coherent optical system over 2000 km SSMF transmission. Insets are the constellation points of (i) QPSK-OFDM, (ii) PCS16-OFDM, and (iii) QPSK-NOFDM signals.}
	\label{2000km_BER}
\end{figure}

Fig. \ref{LOP} shows the BER versus launched optical power (LOP) of the 400G coherent optical system over 2000 km SSMF transmission without ASE noise loading. The QPSK-NOFDM signal also has the advantage of good tolerance to fiber nonlinearity. OSNR increases as the LOP increases, but the fiber nonlinearity also increases. The measured optimal LOP of the QPSK-OFDM, PCS16-OFDM, and QPSK-NOFDM signals is approximately 5 dBm, 4 dBm, and 5 dBm, respectively. BER versus OSNR of the 400G coherent optical system over 2000 km SSMF transmission is shown in Fig. \ref{2000km_BER}. The LOP of the QPSK-OFDM, PCS16-OFDM, and QPSK-NOFDM signals is set to the optimal LOP. Insets are the constellation points of (i) QPSK-OFDM, (ii) PCS16-OFDM, and (iii) QPSK-NOFDM signals at the OSNR of 20 dB. QPSK-OFDM signal cannot reach the 20\% FEC limit. The maximum achievable OSNR of the PCS16-OFDM and QPSK-NOFDM signals is approximately 20 dB and 20.7 dB, respectively. The PCS16-OFDM signal achieves a lower BER among these three signals at the same OSNR, while the QPSK-NOFDM can reach the best BER performance due to the higher LOP.

\setlength\tabcolsep{2.5pt}
\begin{table}[!t]
	\centering
	\begin{threeparttable}[b]
	\renewcommand\arraystretch{1.2}
	\caption{Comparison of QPSK-OFDM, PCS16-OFDM, and QPSK-NOFDM for long-haul coherent optical communications.}
	\begin{tabular}{cccc}
		\toprule[1.5pt]
	     & QPSK-OFDM & PCS16-OFDM & QPSK-NOFDM \\ 
		\midrule[1pt]
		Baud rate&130 Gbaud &113.75 Gbaud&130 Gbaud\vspace{1pt} \\
  Coded bits number&1x &1.4583x &1x\vspace{1pt} \\ 
  ROSNR (OBtB)&20.5 dB&18 dB&17.9 dB\\
  ROSNR (11 WSSs)&/&18.6 dB&19 dB\\
  ROSNR (2000 km)&/&18.5 dB&18.7 dB\\
  \begin{tabular}[c]{@{}c@{}}Achievable OSNR\\(2000 km)\end{tabular} 
  &/&20 dB&20.7 dB\\
		\bottomrule[1.5pt] 
	\end{tabular}
       \footnotesize
       Noted: ROSNR: required OSNR at 20\% FEC limit.
     \label{tab}
     \end{threeparttable}
\end{table}

Table \ref{tab} shows the comparison of QPSK-OFDM, PCS16-OFDM, and QPSK-NOFDM for long-haul coherent optical communications. The required OSNR (ROSNR) of the QPSK-OFDM for the OBtB transmission is about 20.5 dB, while it can not reach the 20\% FEC limit over 11 WSSs or the 2000 km transmission due to performance deterioration of the high-frequency subcarriers. PCS16-OFDM has the advantages of a low baud rate and the best WSS filtering tolerance. However, the coded bits of PCS16-OFDM are 1.4583 times that of QPSK-OFDM due to the redundancy caused by the distribution matcher. It also achieves a lower ROSNR over 2000 km transmission. The QPSK-NOFDM shows the comparable WSS filtering tolerance to PCS16-OFDM and superior nonlinearity tolerance with the highest achievable OSNR over a 2000 km transmission link. Moreover, the coded bits of the QPSK-NOFDM are the same as those of the QPSK-OFDM.

\section{Conclusion}
\label{CONCLUSION}
In this paper, we propose the FTN-NOFDM to improve SE for long-haul coherent optical communications. The subcarrier number is set to eight to enable low-complexity FTN-NOFDM signal generation using a pruned IFFT and ICI cancellation. A frequency tone-based TR is proposed for FTN-NOFDM. A TD-MIMO equalizer is designed that updates the tap coefficients based on the outputs of conventional ID. To further mitigate ICI, the LDPC-assisted ID is integrated into the conventional ID module. FTN-NOFDM, PCS-OFDM, and the QPSK-OFDM are experimentally compared in a 400G coherent optical communications system over 11 cascaded WSSs and 2000 km transmission. The results show that the FTN-NOFDM exhibits comparable filtering tolerance to PCS-OFDM and superior nonlinearity tolerance. Moreover, it can be expected that the WSS filtering tolerance can be enhanced by increasing the subcarrier number of FTN-NOFDM at the expense of increasing computational complexity. In conclusion, the proposed FTN-NOFDM shows potential in improving SE for 400G and beyond optical transport networks.


\bibliographystyle{IEEEtran}
\bibliography{sample}

\vfill

\end{document}